\documentclass[preprint2]{aastex61}
\usepackage{rotating}

\begin{document}

\title{\replaced{Direct Measurements of the Composition of Exoplanet Interiors with the James Webb Space Telescope}{Inferring the Composition of Disintegrating Planet Interiors from Dust Tails with Future James Webb Space Telescope Observations} }

\author[0000-0002-4133-5216]{Eva H.\ L.\ Bodman}
\altaffiliation{NASA Postdoctoral Program Fellow, Nexus for Exoplanet System Science}
\affiliation{School of Earth and Space Exploration, Arizona State University, P.O. Box 871404, Tempe, AZ 85287-1404, USA}

\author[0000-0001-6160-5888]{Jason T.\ Wright}
\affiliation{Department of Astronomy \& Astrophysics, The Pennsylvania State University, 
525 Davey Laboratory,
University Park, PA, 16802, USA}
\affiliation{Center for Exoplanets and Habitable Worlds, The Pennsylvania State University, 
525 Davey Laboratory,
University Park, PA, 16802, USA}
%\affiliation{NASA Nexus for Exoplanet System Science}

\author{Steven J.\ Desch}
\affiliation{School of Earth and Space Exploration, Arizona State University, P.O. Box 871404, Tempe, AZ 85287-1404, USA}
%\affiliation{NASA Nexus for Exoplanet System Science}

%\author{Ming Zhao}
%\altaffiliation{Now at the New York Times Company}
%\affiliation{Department of Astronomy \& Astrophysics, The Pennsylvania State University, 525 Davey Laboratory, University Park, PA, 16802, USA}
%\affiliation{Center for Exoplanets and Habitable Worlds, The Pennsylvania State University, 525 Davey Laboratory,University Park, PA, 16802, USA}
%\affiliation{NASA Nexus for Exoplanet System Science}

\author{Carey M.\ Lisse}
%\affiliation{NASA Nexus for Exoplanet System Science}
\affiliation{JHU-APL, 11100 Johns Hopkins Road, Laurel, MD 20723, USA}

%\author{Daniel Jontof-Hutter}
%\affiliation{University of the Pacific, 3601 Pacific Avenue, Stockton, CA 95211, USA}
	
%\author{Natasha E.\ Batalha}
%\altaffiliation{Need her new affiliation}
%\affiliation{Department of Geosciences, 503 Deike Building, The Pennsylvania State University, University Park, PA, 16802, USA}
%\affiliation{Department of Astronomy \& Astrophysics and Center for Exoplanets and Habitable Worlds 525 Davey Laboratory The Pennsylvania State University University Park, PA, 16802, USA}

\begin{abstract}
Disintegrating planets allow for the unique opportunity to study the composition of the interiors of \added{small, hot, rocky} exoplanets because the interior is evaporating and that material is condensing into dust, which is being blown away and then transiting the star. Their transit signal is dominated by dusty effluents forming a comet-like tail trailing the host planet (or leading it, in the case of K2-22\added{b}), making these good candidates for transmission spectroscopy. To assess the ability of such observations to diagnose the dust composition, we simulate the transmission spectra \added{from 5-14 $\mu$m } for the planet tail assuming an optically-thin dust cloud comprising a single dust species\added{ with a constant column density scaled to yield a chosen visible transit depth}. We find that silicate resonant features near 10 $\mu$m can produce transit depths that are at least as large as those in the visible. For the average transit depth of 0.55\% in the \textit{Kepler} band for K2-22\added{b}, the features in the transmission spectra can be as large as 1\%, which is detectable with the \textit{JWST} MIRI low-resolution spectrograph \added{in a single transit}. The detectability of compositional features is easier with an average grain size of 1 $\mu$m despite features being more prominent with smaller grain sizes. We find most features are still detectable for transit depths of $\sim 0.3$\% in the visible range. If more disintegrating planets are found with future missions such as the space telescope \textit{TESS}, follow-up observations with \textit{JWST} can explore the range of planetary compositions.
\end{abstract}


\section{Introduction}

During the Kepler mission \citep{Borucki09}, a new subset of the ultra-short period planets \citep[``USP planets,'' e.g.][]{Sanchis14}, referred to as ``disintegrating planets" was discovered. These appear to have large dusty tails as a result of their disintegration, presumably due to the intense irradiation of their host star \citep[e.g.\ KIC 12557548,][]{Rappaport12}. The transits of these disintegrating planets are characterized by a sharp and steep ingress followed by an extended tail-like egress \added{or the reverse (extended ingress and steep egress) as in the case of K2-2b}. Some systems also show a short brightening before the ingress and a brightening after the egress, suggestive of forward scattering in a dust tail.  Currently, the most promising hypothesis for these systems is a low-mass (lunar or Mercury-like) disintegrating planet surrounded by escaping dusty effluents that form a comet-like tail much larger ($\sim$10x) than the rocky planet itself \citep{Rappaport12}. The dayside temperatures of these planets ($\sim$2000K) are high enough to evaporate rock and cause large mass loss. The transit depths of these planets are highly variable, \replaced{apparently due to some sort of instability in the dust cloud tails that changes the opacity or orientation on timescales shorter than an orbit.}{possibly due to changes in the dust production rate which may linked to stellar activity \citep{Croll15,Kawahara13}.}

%\textbf{integrate these 2 paragraphs}
 Depending on the mechanism by which the material is leaving its parent body, the dust grains composing the comet-like tail may be a representative sample of the interior of the planet which can be examined through transmission spectroscopy. %  And unlike the bulk elemental information provided by white dwarf pollution, it may retain important mineralogical information about its parent body, including its degree of hydration.  Thus there is the potential in these bodies for us to know {\it more} about some aspects of their interiors than we know about the Earth's!
\deleted{Direct measurement of the chemical composition of planetary interiors is generally difficult to impossible, even for Solar System planets, including the Earth, but would provide crucial information for studies of habitability. A planet is considered to be in the Habitable Zone if its surface temperatures can support the existence of liquid water \citep{Kasting93,Kopparapu13}, but surface habitability of a planet depends on the composition of the planetary interior. For example, plate tectonics, which depends on composition, sustains global cycles of important bioessential elements such as C and P.}
%sensitively on the degree of greenhouse warming in its atmosphere \citep[e.g.][]{Kopparapu14}, which is a strong function of bulk planetary properties, as well as both surface {\it and interior} composition. 
%Carbon dioxide is an excellent greenhouse gas, but whether carbon is emitted from a planet's interior as ${\rm CO}_{2}$ (as in volcanic outgassing) or ${\rm CH}_{4}$ (as from shallower hydrothermal sources) depends on the redox state of the subsurface minerals. The amount of ${\rm CO}_{2}$ in Earth's atmosphere (vs. sediments) is regulated by the carbonate-silicate cycle, in which Ca-bearing silicates in the Earth's crust are weathered to produce ${\rm Ca}^{++}$ cations, which then react with carbonic acid in seawater to produce carbonates [${\rm CaCO}_{3}$]. While this process is believed to act on rocky exoplanets, regulation of ${\rm CO}_{2}$ would not be possible without these sources of cations such as Ca. Beyond these factors, the habitability of a planet also depends on the abundances of bioessential elements--- especially S, P, O, N, C and H, but also trace species like Mg (found in chlorophyll), Fe (found in hemoglobin), Mn (a key element in enzymes used for nitrogen fixation), and a host of others. \textbf{citations? Any other examples of the importance of composition?}
Practitioners of transmission spectroscopy of \deleted{typical} exoplanets typically must deal with a challenging geometry: the planetary transit itself is usually not difficult to detect, but the contribution to the signal from the exoplanetary atmosphere is often orders of magnitude smaller \added{for small, rocky planets}. \deleted{Despite this difficulty,} Transmission spectroscopy has been widely applied to study the atmospheres of exoplanets with large scale heights, such as hot Jupiters, super-Neptunes, and super-Earths. These studies have significantly broadened our knowledge of planetary atmospheres.  In the infrared (IR), molecules that are important to habitability, such as H$_2$O, CH$_4$, CO$_2$, and CO, have been found in some planets using transmission spectroscopy \citep[e.g.,][]{Barman2008, Swain2008,Wakeford2013}. 
\replaced{In contrast, the transit signatures of disintegrating planets appear to be entirely due to the very signal we seek, and so provide an opportunity for relatively high signal-to-noise ratio detections of spectral features without very large allocations of telescope time and a dedicated data reduction effort.
Indeed, the transit depths of KIC 12557548 (up to 1.3\%) % and WD 1145 (up to 40\%)
are larger than those of most exoplanetary transits.}{Unlike most terrestrial exoplanets, the transit signal of disintegrating planets is almost entirely from the dust tail, making these planets strong candidates for transmission spectroscopy, like planets with atmospheres at large scale-heights.}

Currently, three systems \added{around main sequence stars} have been discovered with apparent cometary tails of disintegrated material\replaced{, but t}{: KIC 12557548, KOI 2700, and K2-22. One white dwarf, WD 1145+017, also displays period transits of debris with cometary tails \citep{Vanderburg15}, but the dust tail is likely due to tidal disruption unlike those in the other systems. T}here should be many more such objects in short-period orbits according to thermal evaporation models \citep{Perez13}, and current and future missions like {\it K2} and {\it TESS} are expected to discover more of these objects around nearby bright stars that are well suited for characterization. \citet{DeVore16} argue that there may be even more such planets detected via brightenings due to forward scattering by non-transiting dust.

% \begin{figure}[tbp]
% \centering
% %\mbox{
% \includegraphics[angle=0,width=2.2in,height=1.7in]{Ming_Jason/KIC1255_lc.png}
% %\hspace{0.01in}
% \includegraphics[angle=0,width=2.2in,height=1.7in]{Ming_Jason/WD1145.png}
% %\hspace{0.02in}
% \includegraphics[angle=0,width=2.in,height=1.6in]{Daniel/KIC84lightcurve.pdf}
% %} 
% %\vspace{-0.01in}
% \caption{
% {\footnotesize 
% Transit light curves for KIC 1255b (left), K2-22b (middle) and KIC 8462852 (right). The light curve of KIC 1255b (left) is the average of 4 quarter's of {\it Kepler} data. For K2-22b (middle), the red curve shows the expected transit for a solid body, while the points and blue curve show the transit data and the corresponding model  --  the long, gradually changing egress is a strong signature of a comet-like tail. The right panel shows the deep, unrepeated and asymmetric transit event of KIC 8462852.
% }}
% \label{fig_light_curve}
% \vspace{-0.11in}
% \end{figure}

%\paragraph{Interpretation of the shapes of the light curves}
%	\input{Daniel/intro_Daniel.tex}

%\subsection{Known Disintegrating Objects}
\subsection{KIC 12557548}
KIC 12557548 was the first system discovered to have an orbiting disintegrating planet \citep{Rappaport12}. The planet transits the mid-K dwarf host star (T$_{\rm eff}$= 4300 K) at a short orbital period of 15.7 h and its transit light curve has the distinct asymmetric shape of a comet-like dusty tail. %Today, the consensus scenario is that the observed transits are entirely due to a comet-like dusty tail trailing behind a small planet \citep{Brogi12, Budaj13}. 
The  transit duration is much longer than the crossing time of the planet ($\sim$0.1 of the orbit), suggesting the tail is longer than even the size of the star. The deep transit depth ($<1.3\%$) and the comet-like tail were argued to be from plumes of dusty clouds \replaced{evaporated}{condensed from the rock vapor} from the planet's highly irradiated surface (2200 K) \citep{Budaj13}. 
The \textit{Kepler} light curve of KIC 12557548b also has a prominent pre-transit brightening and a weaker post-transit brightening, suggesting forward scattering by particles of radius of 0.1 - 1 \micron\ in the dusty tail \citep{Budaj13}. \citet{Schlawin16} finds evidence for large particles in the tail from the flatness of the transmission spectrum, but this data was taken after the \textit{Kepler} data during a smaller-than-average transit, which may indicate the system was in a weak period or the mechanism has changed. 

The size of the dusty cloud, inferred from the transit depths, suggests a small rocky planet (based on its Hill radius) that has a rough mass loss rate of 1 M$_\Earth$Gyr$^{-1}$ \citep{Rappaport12}. Since it is extremely difficult on theoretical grounds to remove dust from large planets with substantial surface gravity, \citet{Perez13} concluded that KIC 12557548b is lower in mass than Mercury, and if the mass loss rate estimate is correct then the planet may be as small as 2 lunar masses, and may have lost $\sim$70\% of its original mass, implying that it is in the final catastrophic disintegrating phase. This makes KIC 12557548b one of the smallest known exoplanets. (The smallest is Kepler 37b, \citealt{Barclay2013}.)

KIC 12557548's transit depth varies stochastically on timescales shorter than the orbital period, which might be caused by eruptions of dust (as in a cometary nucleus), variations in stellar radiation to the planet's surface, and/or volcanic activities at the planet's surface. All these hypotheses require more detailed theoretical and quantitative modeling to assess their credibility, and likely will imply different transmission properties of the cloud. 
There is evidence that its transit depth correlates with stellar rotation, likely due to the planet occulting star spots rather than enhanced mass loss \citep{Croll15}. 
The variability of the optical depth of the dusty cloud implies changes in the transit shape as well, but such variation is too small to detect for KIC 1255b due to the faintness of its host star ($K_p=15.7$ mag).% \textbf{Is this paragraph needed?}

%Long term quiescent period?
%open questions?
%Determining the size and mineralogy of this dust would inform models of this process and probe the chemistry of such exoplanets' crusts. 
\begin{deluxetable}{lccc}
\tabletypesize{\scriptsize}
%\rotate
\tablecaption{System Properties \label{system_prop}}
\tablewidth{0pt}
\tablecolumns{4}
\tablehead{
\colhead{Property (Units)}  & \colhead{KIC 1255} &  \colhead{KOI 2700} & \colhead{K2-22}
    }
\startdata
Host Star Properties &	&	&	\\
\hline
Magnitude $J$ (mag)				& 14.02	& 13.58	& 12.74	\\
Stellar Temperature (K)			& 4500	& 4300	& 3830	\\
Surface gravity, log$g$ (cgs)	& 4.62	& 4.71	& 4.65	\\
Metallicity [Fe/H] 				& -0.2	& -0.7	& 0.03	\\
\hline
Planet Transit Properties	&	&	&	\\
\hline
Orbital period (hr)	& 15.68	& 21.84	& 9.146	\\
Transit length (hr)		& 1.5	& $>5$	& 0.8	\\
Transit depth range (\%) & 0-1.4	& 0.03-0.053	& 0-1.3	\\
Mean transit depth (\%)	& 0.5	& 0.036	& 0.55	\\
\enddata
%\tablecomments{citations?}
\tablerefs{\citet{Rappaport12} for KIC 1255; \citet{Rappaport14} for KOI 2700; \citet{Sanchis-Ojeda2015}}
%\tablenotetext{a}{notes}
\label{tab:sysprop}
\end{deluxetable}

\subsection{KOI 2700} 
KOI 2700b is reminiscent of KIC 12557548b in that its light curve shows a distinct asymmetric transit curve with a sharp ingress and a shallow egress, likely caused by a trailing dusty comet-like tail \citep{Rappaport14}. Its average transit depth decreased by a factor of 2 over the course of the $Kepler$ mission ($\sim$4 years), indicating a decrease of mass-loss rate or precession of the orbit out of the line of sight.  The egress of the transit lasts for $\sim$25\% of the orbital period (21.84 hr), indicating a long lifetime for the dust grains and the tail. The transit depth of KOI 2700b ($\sim$360 ppm) is very shallow, suggesting a mass-loss rate an order of magnitude smaller than that of KIC 12557548b, and is too small to allow for analysis of transit-to-transit variation as with KIC 12557548b. The corresponding rocky core is expected to be as small as the Moon \citep{Rappaport14}.  The shallow and decreasing transit depth of KOI 2700b ($\sim$360 ppm) makes followup difficult; however, if the trend in depths reverses and the transits become deep again, it may make an excellent target.

\subsection{K2-22} 

Another system that shows asymmetric transit light curves is the K2-22 system \citep{Sanchis-Ojeda2015}, which includes a planet with a very short period of only 9.1457 hours around an M-dwarf host star (T$_{\rm eff}\sim3800$ K). Unlike KIC 12557548, the transit light curve of K2-22 does not have a significantly shallower egress. Instead, it shows a prominent post-transit brightening bump and a weaker pre-transit brightening, both of which are better interpreted as forward scattering by a dominant leading dusty tail and a trailing tail with a large scale length of almost half of its host star's radius \citep{Sanchis-Ojeda2015}. 
 Like KIC 12557548b, the transit depths are highly variable, ranging from $\leq$0.14\% to 1.3\%. The shape of its transit light curves also appears to be variable. All these features point to dusty effluents escaped from the surface of a small rocky planet within a Moon-to-Mercury mass range.
 
 Although the disintegrating planet hypothesis succeeds in explaining ground- and space-based observations, the particle sizes predicted by the two-dusty-tail model for K2-22b are not entirely self-consistent \citep{Sanchis-Ojeda2015}. In addition to grains being launched forwards along the orbit, the formation of a leading tail requires the ratio of radiation pressure forces to gravity to be low. This occurs for a low-luminosity host star for very small ($\lesssim0.1\,\mu$m) or very large grains ($\gtrsim1\, \mu$m). However, \added{\citet{Sanchis-Ojeda2015} found that the} best fits of forward scattering bumps in the light curve prefer a grain size of $\sim0.5\,\mu$m\replaced{, as do those for the color dependence of the transits}{. The authors also found a weak color dependence in the transit depth for one of the dips they observed with ground-based spectroscopy which indicates dust with a grain size $\gtrsim0.5\,\mu$m.} Significant theoretical work is therefore still needed to reconcile all the observations and predictions. % \textbf{This should probable be expanded since we are focusing on this one.}
%variations in the transit shapes; and demonstrate clearly that at least on one occasion the transit depths were significantly wavelength dependent.
Since K2-22 is brighter than KIC 12557548 ($K_p=14.9$), this star is currently the best target for follow-up observations, assuming the transit depths continue to exceed 1\%. A summary of these planetary system properties is listed in table \ref{tab:sysprop}.

In this paper, we examine the detectability of the spectroscopic signatures  of the composition of the dust grains in the disintegrating planet tails, in particular with regard to the capabilities of the upcoming {\it James Webb Space Telescope} ({\it JWST}) mission.  
%The rest of this paper is organized as follows.
 In section \ref{sec:mineralogy}, we discuss the \deleted{expected internal mineralogy of a terrestrial planet and }composition of the dust that condenses out of vaporized rock material, assuming Earth-like composition. 
Then, in section \ref{sec:spectroscopy}, we describe our simulated transmission spectra from different possible dust species that may be produced from the disintegration of interior planetary material. In section \ref{sec:JWST}, we use our simulated spectra to estimate the detectability of compositional features with \textit{JWST}. %Finally, we summarize our findings in section \ref{sec:conclusions}. 

\section{Exoplanetary mineralogy} \label{sec:mineralogy}

Disintegrating planets offer a tantalizing opportunity to measure exoplanetary interior composition directly. While the exact mechanism of producing the dust cloud is uncertain, the material is expected to be representative of an actual exoplanetary interior. The structure of the tail offers some compositional clues. Based on the lengths of tails and estimated dust grain lifetimes, and the sublimation properties of various minerals, \citet{vanLieshout2014} concluded that corundum (Al$_2$O$_3$) and iron-rich silicates could be major components in the tail of KOI 2700b and of the minerals considered, only corundum was consistent with KIC 12557548b \citep{vanLieshout2016}. However, spectral information are more powerful. Since the dust cloud obscures the starlight at optical wavelengths, rudimentary characterization of composition can be done from the light curve transits alone \citep{vanLieshout2014, vanLieshout2016}. Supplementary IR spectra of disintegrating planet systems can constrain the dust compositions even more powerfully.

To assess the likely mineralogy of the dust, we adopt the hypothesis that the dust seen in tails that transit stars in disintegrating planet systems arises from the planets as follows. The planets are close enough to their host stars that their surface materials should be vaporizing. For the rocky materials that are expected to dominate, Fe,Mg-rich silicate minerals like olivine ($({\rm Mg,Fe})_2 {\rm SiO}_4$) or pyroxene ($({\rm Mg,Fe}){\rm SiO}_3$), this requires temperatures above about 1200 K at $P \sim 10^{-6}$ atm,
to 1450 K for $P > 10^{-3}$ atm \citep{Ebel2000, Ebel2006}. 
Ca,Al-rich silicates (e.g., hibonite, perovsite, grossite, melilite) and corundum (${\rm Al}_{2}{\rm O}_{3}$) vaporize at higher temperatures: 1450 K at $P \sim 10^{-6}$ atm, to 2000 K at $P \sim 1$ atm \citep{Ebel2000, Ebel2006}. 
No abundant minerals (based on cosmochemical/planetary abundances) are expected to survive at temperatures greater than about 2000 K. A disintegrating planet with surface temperatures at or above $\sim$1200 K can build up an atmosphere of rock vapor that, if not gravitationally bound to the planet, will escape into space. As it escapes, the rock vapor expands and cools, allowing condensation of minerals from the vapor that evaporated. Whether the rock vapor condenses in the atmosphere or in the dust tail and the exact composition of the condensates is uncertain. Silicate clouds can form above magma oceans on Super-Earths assuming an adiabatic pressure profile \citep{Schaefer2009} but detailed modeling of disintegrating planet atmospheres and wind is needed.

In analogy to other condensates seen in other astrophysical systems, the grains that condense very likely will be micron-sized. Amoeboid olivine aggregates (AOAs) in chondritic meteorites consist of grains $5 - 20 \, \mu{\rm m}$ in diameter that condensed from the solar nebula \citep{Scott2014}. Models of dust condensation in asymptotic-giant-branch (AGB) outflows suggest grain growth to a maximum size of $\sim 1 \, \mu{\rm m}$ \citep{Hofner1996}. The average radius of grains condensed from the winds of the red supergiant VY CMa is observationally constrained to be $\approx 0.5 \, \mu{\rm m}$ \citep{Scicluna2015}. Observations of dust condensed out of the Crab Nebula (core-collapse) supernova remnant likewise favor dust sizes just under $\sim 1 \, \mu{\rm m}$ in radius \citep{Owen2015}. The maximum radii of dust grains in the interstellar medium are $\sim 0.25 \, \mu{\rm m}$ \citep{Mathis1977}. If our hypothesis that the dust from disintegrating planets is condensed from their rock vapor outflows, then the optical or, preferably, mid-infrared transmission spectra of these micron-sized dust grains
can provide information on the stoichiometry of the vapor from which they condensed, and therefore the planets from which they are derived.

The minerals that would condense from these starting compositions would depend on the uncertain temperature to which the planet's surface is raised. If temperatures are not high enough to vaporize Ca or Al, they would not be available to condense into minerals. Presence or absence of Ca or Al is more a function of temperature than intrinsic Ca or Al to Si ratios. However, some elements have similar condensation/vaporization temperatures and the ratio of those elements in condensed minerals would be diagnostic of the ratios in the evaporating surface material. In particular, the 50\% condensation temperatures of Fe, Mg, and Si (in a solar-composition gas at $10^{-4}$ atm pressure) are 1328 K, 1327 K, and 1302 K, respectively; for comparison, those of Ca and Al are 1505 K and 1641 K, respectively, and those of Na and K are  953 K and 1001 K, respectively \citep{Lodders2003}. Fe, Mg, and Si are likely to vaporize together and condense together, with the elemental ratios in the dust grains reflecting the molar ratios in the current surface material.

In particular, it may be possible to distinguish whether dust is condensing from material derived from the crust, mantle (assuming the crust has been lost already), or core (assuming the mantle has been lost already) of a planet. \added{For example,} the Earth's continental crust is stoichiometrically \replaced{60wt\% ${\rm SiO}_{2}$ (silica), 15wt\% ${\rm Al}_{2}{\rm O}_{3}$ (alumina), 3.8wt\% FeO (Fe {\sc ii} oxide) and 2.5wt\% ${\rm Fe}_{2}{\rm O}_{3}$ (Fe {\sc iii} oxide), 5.5wt\% CaO, 3.1wt\% MgO, 3.0wt\% ${\rm Na}_{2}{\rm O}$, 2.8wt\% ${\rm K}_{2}{\rm O}$, 1.4wt\% ${\rm H}_{2}{\rm O}$, 1.2wt\% ${\rm CO}_{2}$ (in carbonates), 0.7wt\% ${\rm TiO}_{2}$, and sub-percent fractions of other elements \citep{Brown1981}.}{majority silica (60wt\% ${\rm SiO}_{2}$). Crustal material also includes the minerals listed with wt\% in table \ref{tab:Earth_Comp} along with sub-percent fractions of other elements \citep{Brown1981}.} 
 The Earth's oceanic crust is similar but distinct, being stoichiometrically \replaced{49wt\% ${\rm SiO}_{2}$, 17wt\% ${\rm Al}_{2}{\rm O}_{3}$, 12wt\% CaO, 6.2wt\% FeO and 2.3wt\% ${\rm Fe}_{2}{\rm O}_{3}$, 6.8wt\% MgO, 2.6wt\% ${\rm Na}_{2}{\rm O}$, 0.4wt\% ${\rm K}_{2}{\rm O}$, 1.1wt\% ${\rm H}_{2}{\rm O}$, 1.4wt\% ${\rm CO}_{2}$ (in carbonates), 1.4wt\% ${\rm TiO}_{2}$, etc.}{less silica and more FeO, CaO, and MgO} \citep{Brown1981}. These crustal compositions contrast strongly with the composition of the Earth's mantle, which stoichiometrically \replaced{45wt\% ${\rm SiO}_{2}$, 38wt\% MgO, 8wt\% FeO, 4wt\% ${\rm Al}_{2}{\rm O}_{3}$, 3.5wt\% CaO, 
0.3wt\% ${\rm Na}_{2}{\rm O}$, 0.2wt\% ${\rm TiO}_{2}$, 0.25wt\% NiO, 0.03wt\% ${\rm K}_{2}{\rm O}$,etc.}{has much more MgO, more FeO, and less Al$_2$O$_3$} \citep{McDonough1995}. \added{A more detailed comparison of continental crust, oceanic crust and mantle composition is listed in table \ref{tab:Earth_Comp}.} In further contrast, the Earth's core is predominantly Fe metal ($\approx 86$wt\%) and Ni metal ($\approx 4$wt\%), with light alloying elements of uncertain type and amounts, but arguably $\approx 6$ wt\% Si, $\approx 3$wt\% O, and $\approx 1$wt\% S \citep{Hirose2013}.

\begin{deluxetable}{lccc}
\tabletypesize{\scriptsize}
%\rotate
\tablecaption{Earth Composition \label{tab:Earth_Comp}}
\tablewidth{0pt}
\tablecolumns{4}
\tablehead{
\colhead{Mineral}  & \colhead{Cont. Crust} &  \colhead{Ocean Crust} & \colhead{Mantle}
    }
\startdata
SiO$_2$			& 60	& 49	& 45	\\
Al$_2$O$_3$		& 15	& 17	& 4	\\
FeO				& 3.8	& 6.2	& 8	\\
Fe$_2$O$_3$		& 2.5	& 2.3	& 	\\
CaO				& 5.5	& 12	& 3.5	\\
MgO				& 3.1	& 6.8	& 38	\\
Na$_2$O			& 3.0	& 2.6	& 0.3	\\
K$_2$O			& 2.8	& 0.4	& 0.03	\\
H$_2$O			& 1.4	& 1.1	&	\\
CO$_2\,^a$			& 1.2	& 1.4	&	\\
TiO$_2$			& 0.7	& 1.4	& 0.2	\\
\enddata
%\tablecomments{citations?}
\tablerefs{\citet{Brown1981} for crustal abundances; \citet{McDonough1995} for mantle abundances}
\tablenotetext{a}{in carbonates}
\label{tab:sysprop}
\end{deluxetable}

The ratio of Fe and Mg atoms to Si atoms in the vapor determines whether the main condensates will be silica and pyroxenes [ $({\rm Fe}+{\rm Mg}) / {\rm Si} < 1$ ],
pyroxenes and olivines [ $1 < ({\rm Fe}+{\rm Mg}) / {\rm Si} < 2$ ], or olivines and periclase/w\"{u}stite (MgO and FeO) [ $2 < ({\rm Fe}+{\rm Mg}) / {\rm Si}$ ].
These minerals are easily distinguished in the mid-infrared \citep{Molster2005, Wooden2007, Henning2010}, so IR observations should be able to constrain this elemental ratio.  \added{For an Earth-like composition,} vaporized continental crust, with $({\rm Fe}+{\rm Mg}) / {\rm Si} \approx 0.08$, should recondense overwhelmingly as silica, with some pyroxenes. Vaporized oceanic crust, with $({\rm Fe}+{\rm Mg}) / {\rm Si} \approx 0.33$, should recondense predominantly as silica, but with a higher proportion of pyroxenes. Vaporized mantle, with $({\rm Fe}+{\rm Mg}) / {\rm Si} \approx 1.42$, in contrast, should condense as a mixture of pyroxenes and olivines and not silica. Therefore the presence of silica may be diagnostic of crustal source material, while olivines may be diagnostic of mantle source material. \added{Planet models are needed to test how diagnostic the $({\rm Fe}+{\rm Mg}) / {\rm Si}$ ratio is for determining crust versus mantle material when the composition is significantly not Earth-like. }

The Fe/(Mg+Fe) ratio in silicates is easily observable through the shift in the $10 \, \mu{\rm m}$ silicate absorption feature it creates \citep[e.g.,][]{Molster2005, Wooden2007, Henning2010}, and the presence of Fe in silicates can also be inferred through the $0.9 \, \mu{\rm m}$ absorption feature \citep{Cloutis1991}. 
% continental crust
% 3.8wt\% FeO (72) +2.5wt\% Fe2O3 (160) +3.1wt\% MgO (40) + 60wt\% SiO2 (60)
%  0.0528          + 0.0313 = 0.0841, vs.  0.0775  vs.  1.00 
% Fe / (Fe+Mg) = 0.52 ,    (Fe+Mg) / Si = 0.0841
% oceanic crust
% 6.2wt\% FeO (72) +2.3wt\% Fe2O3 (160) +6.8wt\% MgO (40) +49wt\% SiO2 (60)
%  0.0861          + 0.0144 = 0.1005, vs.  0.170  vs. 0.817
% Fe / (Fe+Mg) = 0.37 
% mantle
% 8.0wt\% FeO (72) +0.0wt\% Fe2O3 (160) +38wt\% MgO (40) +45wt\% SiO2 (60)
%  0.111                              vs. 0.95    vs. 0.75 
% Fe / (Fe+Mg) = 0.11 
% 
Silicates condensed from the Earth's continental crust, oceanic crust, and the mantle, would have molar ratios ${\rm Fe} / ({\rm Mg} + {\rm Fe}) \approx $ 0.52, 0.37, and 0.11, respectively. Dust condensed from the outflow of an evaporating core would condense predominantly Fe metal,
but to the extent that silicates formed, they would lack Mg altogether. Given that an disintegrating exoplanet may have had a very different starting composition from the Earth, it may be difficult to unambiguously identify a molar ratio with continental or oceanic-like crust, but the differences in composition between crust, mantle, and core are stark and potentially identifiable. The exact ratio could provide information about the initial Fe/Mg ratio in the planet or the redox state of the mantle. 

These examples involving Fe, Mg, and Si demonstrate just some of the power of astromineralogical observations of disintegrating planet systems to derive information about planetary compositions. Much more information is potentially derivable through observations that identify Ca- or Al-bearing minerals, or metal grains, etc.

\section{Detectability with Transmission Spectroscopy} \label{sec:spectroscopy}
 % Introduction to transmission spectroscopy of exoplanet atmospheres.

%{\em Ming: 0.8 page + 1 figure}:
%A brief summary of the current status of ground and space-based spectroscopic observations of exoplanetary atmospheres: photometry and spectroscopy, wavelength ranges, spectral resolution, and precision achieved. A few examples of our observations, including the one for KIC 1255. 

Both ground- and space-based resources will need to be brought to bear on the problem.  Ground-based telescopes will be necessary for both their wavelength coverage (complementing the IR coverage of {\it JWST}) and for their monitoring capabilities. % The broad wavelength coverage will be essential to get a complete picture of the dynamics, distribution, and mineralogy of disintegrating planets. 
%The broadband slope from optical to NIR tells the size distribution of scattering dust particles; FeO has an important signature at $\sim$900 nm; CO, H$_2$O have dominating bands in the near-IR; and minerals tracing a planet's crust and interior such as corundum, olivines, and pyroxenes dominate the MIR.   
% Ground-based facilities not only complement {\it JWST} in wavelength, but also provide large aperture and long-term monitoring capabilities. These properties are ideal for observing the faint M-dwarf hosts of most disintegrating planets and for understanding the time variability of their transits, enabling detailed preparatory study for $JWST$.  They will also allow optimization of observing time for the much more expensive and limited space observing time of both $HST$ and $JWST$ (e.g., trigger observations only when the disintegrating planets are in active mass loss phases). 
One major difficulty with observing disintegrating planets is, and will continue to be, their variable depths. Measurements of spectral slopes and features across different instruments must be made simultaneously, lest variations in a transit depth with time be mistaken for those with wavelength.  Spaceborne telescope time is precious and difficult to schedule, which will make it difficult to measure deep transits on temperamental sources.  

New ground-based and existing {\it Kepler} and {\it K2} light measurements can yield statistics about the frequency of deep transits, since in many cases systems seem to go through active and quiescent phases.  It may be possible to predict that at least one of, say, three consecutive transits during an outburst will be deep enough to achieve one's science goals, with some confidence.  If active phases persist long enough, monitoring by more easily-scheduled ground- and space-based telescopes may allow for {\it JWST} to be triggered for targets of opportunity. Fortunately, the nature of USP planets is that over the course of a small number of days a few or several transits can be observed, increasing the likelihood that a sufficiently deep transit will be observed if conditions are favorable. Since the transit depth reaches over 1\% and it is the brightest of the three, K2-22 is currently the best target for follow-up. The planet's 9 hour orbit also allows for quick establishment of its ephemeris and transit depth statistics.

\begin{deluxetable}{lcc}
\tabletypesize{\scriptsize}
%\rotate
\tablecaption{Optical properties of minerals\label{mineral_list}}
\tablewidth{0pt}
\tablecolumns{3}
\tablehead{
\colhead{Mineral}  & \colhead{Wavelength Range} &  \colhead{Ref} \\
\colhead{}  & \colhead{$\micron$} &  \colhead{}
    }
\startdata
 Fe (metallic iron)					& 0.7-200	& O88		\\
 C (carbonaceous dust)    			& .2-800	& D84		\\
 SiO$_2$ (quartz) 					& 6-10000 	& Z13		\\
 MgSiO$_3$ (cryst. enstatite) 		& 2-98 		& J98 		\\
Mg$_2$SiO$_4$ (cryst. forsterite)   & 0.2-2, 2-8000	& F01,Z11	\\
Fe$_2$SiO$_4$ (cryst. fayalite)     & 0.4-10, 2-10000 & F01		\\
Al$_2$O$_3$ (corundum)    			& 0.5-400	& K95		\\
\enddata
%\tablecomments{citations?}
\tablerefs{D84: \citet{Draine1984};F01: \citet{Fabien2001}; J98: \citet{Jaeger1998}; K95:\citet{Koike1995}; O88: \citet{Ordal1988}; Z11: \citet{Zeidler2011}; Z13: \citet{Zeidler2013}}
%\tablenotetext{a}{notes}
\end{deluxetable}
Here, we consider seven dust species (listed in table \ref{mineral_list}): Fe and C, two species that primarily contribute to the spectral continuum; quartz (SiO$_2$); enstatite (MgSiO$_3$), forsterite (Mg$_2$SiO$_4$), and fayalite (Fe$_2$SiO$_4$), three silicates; and corundum (Al$_2$O$_3$), a likely marker for crustal material. Only forsterite and fayalite are included from the olivine mineral family because those two minerals are the end-members, containing only Mg or Fe. Typical olivine is a mixture of these two minerals and displays a blend of the spectral features from those two minerals. The tail will more likely contain a mixture of both forsterite and fayalite. We did not include any iron-rich pyroxenes, as condensation of enstatite over ferrosilite (FeSiO$_3$) is highly favored, and Fe is overwhelmingly seen in solar system silicates as fayalite.

A spherical grain with radius $r$ has an extinction cross section $\sigma(\lambda,r) = \pi r^2 \,Q_\mathrm{ext}$, where $Q_\mathrm{ext}(\lambda,r)$ in the monochromatic extinction efficiency. We calculate the extinction efficiencies using Mie theory and the references for the index of refraction data for each dust species is listed in table \ref{mineral_list}. The index of refraction was extrapolated to shorter wavelengths as needed\added{, see the wavelength range of the data in table \ref{mineral_list}}. We assume a simple power-law grain size distribution, $dn=r^{-a}dr$ with an effective grain radius of $r_\mathrm{eff}=\int r^{1-a}dr/\int r^{-a}dr$.
We average over the grain size distribution to calculate an effective monochromatic extinction cross section,
\begin{equation}
\bar{\sigma}(\lambda)_\mathrm{eff} = \frac{\pi \int_{0.01}^{r_\mathrm{max}} Q_\mathrm{ext}(\lambda,r) r^{2-a}dr}{\int_{0.01}^{r_\mathrm{max}} r^{-a}dr} .
\end{equation}

Since the grain size in the planet tails has been constrained from forward scattering and wavelength dependency of the transit depths to 0.1 - 1.0 $\mu$m \citep{Budaj13, Rappaport14, Sanchis-Ojeda2015}, we consider two size distributions to approximate the observed upper and lower limits of the effective dust grain size. For the small grain size limit, we set the maximum radius ($r_\mathrm{max}$) at 2.0 $\mu$m and set the exponent for each species such that $r_\mathrm{eff}$=0.1 $\mu$m; for the large grain size limit, $r_\mathrm{max}$=5.0 $\mu$m and $r_\mathrm{eff}$=1.0 $\mu$m.

\begin{figure*}
\centering
\mbox{
\includegraphics[width=7.0in]{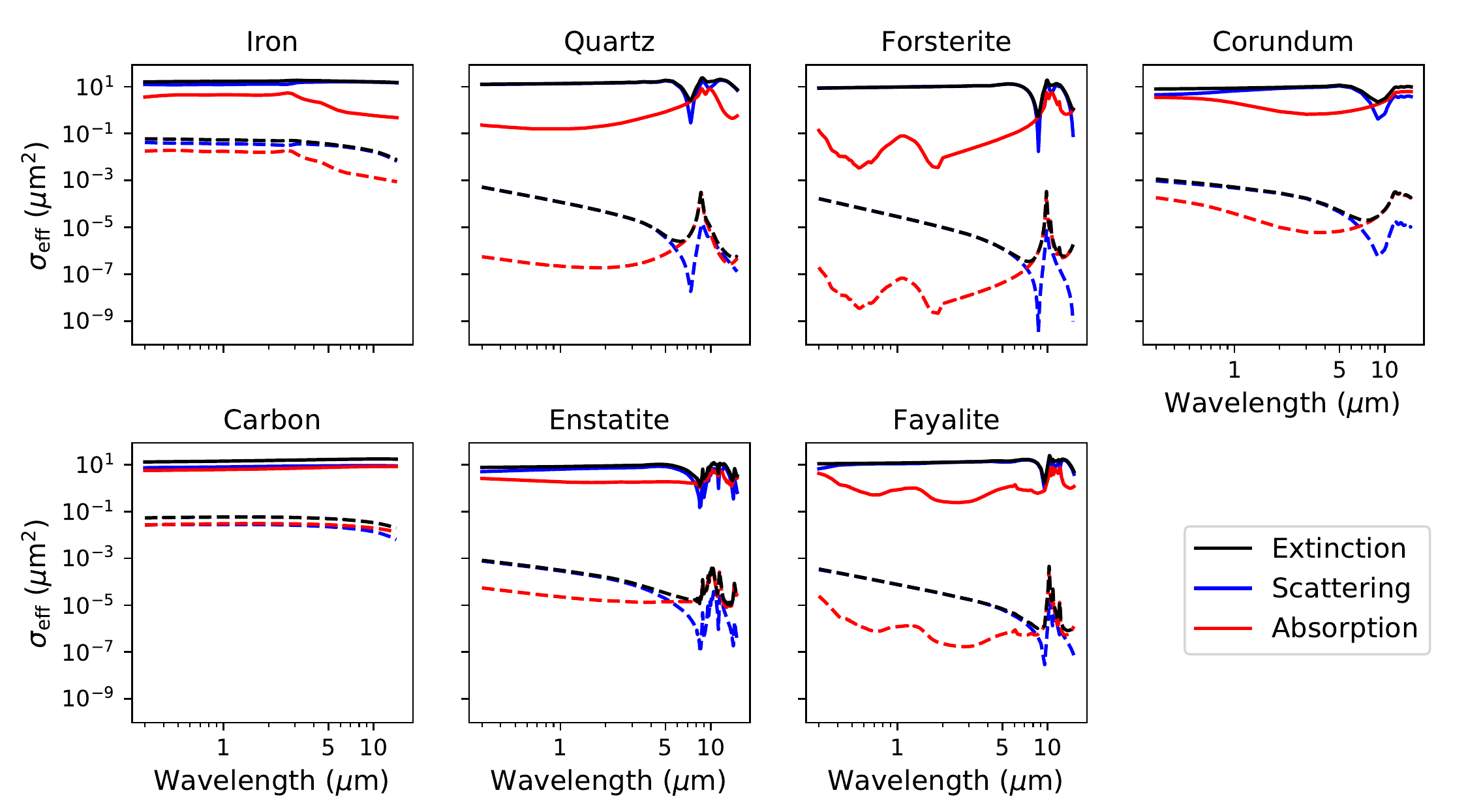}
}
\caption{Extinction, scattering, and absorption cross sections for each of the dust species in black, blue and red respectively. The solid lines display the cross sections for the large grain size distribution and the dashed lines show the small grain size distribution. Scattering dominates at the shorter wavelengths, resulting in a smooth power law dependence that has a size and composition degeneracy that is particularly problematic for the larger grain sizes. Unique compositional features appear at wavelengths longward of $\sim 7\,\mu$m for the silicates and corundum.}
\label{fig:Qs}
\end{figure*}

The effective total extinction cross sections for each size distribution are shown in Fig. \ref{fig:Qs} along with the effective scattering and absorption cross sections which are related by $\bar{\sigma}_\mathrm{ext}=\bar{\sigma}_\mathrm{sca}+\bar{\sigma}_\mathrm{abs}$. For wavelengths shorter than $\sim 6\, \mu$m, the extinction cross section is dominated by scattering, resulting in a featureless curve that depends on the grain size with a degeneracy with composition and cannot be distinguished in this wavelength range alone. 
% The wavelength dependence of the transit reddening does vary with composition but the degeneracy with grain size cannot be distinguished. 
\added{In the 3-7 $\mu$m region, the extinction cross section rapidly decreases with the scattering efficiency for all compositions except iron and carbon, which remain approximately flat to $\sim$15 $\mu$m.}
 Longward of $\sim7$ $\mu$m, silicates have distinguishing spectral features, as does corundum at $\sim10\,\mu$m. Since the unique compositional features occur in the mid-IR, we focus the rest of our analysis over the wavelength range for the \textit{JWST} MIRI spectrograph ($\sim5-14\, \mu$m).

\begin{figure*}
\centering
\mbox{
\includegraphics[width=6.5in]{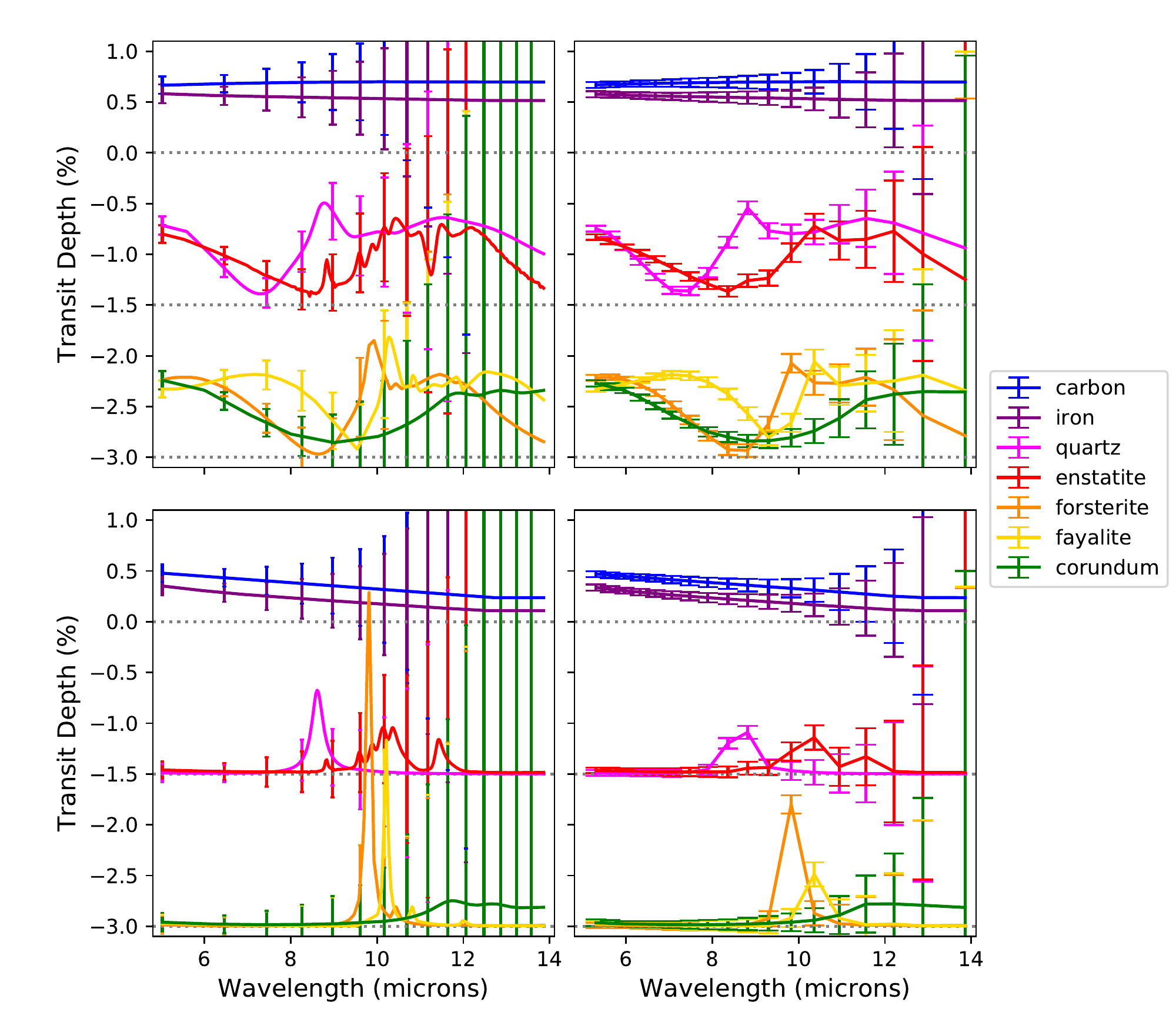}
}
\caption{Model spectra of a single transit assuming the average transit depth for K2-22b (0.55\%) in the visible for expected dust species (listed in Table \ref{mineral_list}) with the two size distributions. The large grain sizes ($r_\mathrm{eff}=1.0\,\mu$m) are in the upper plots and the small grain sizes ($r_\mathrm{eff}=0.1\,\mu$m) are in the lower plots. The left plots are the unbinned spectra with only every 35th error bar shown and the right plots are binned to R=10. %The right plots also include random variance in the data. 
The plots do not include random variance so that the spectral features are more easily seen and the depths are offset for clarity with dotted line marking 0 transit depths for that offset. Colors mark the different dust species: blue for iron, purple for carbon, pink for quartz (SiO$_2$), red for enstatite (MgSiO$_3$), orange for forsterite (Mg$_2$SiO$_4$), yellow for fayalite (Fe$_2$SiO$_4$), and green for corundum (Al$_2$O$_3$). 
\replaced{Size has a large impact on the spectra, but for both distributions, the silicate features near 9 $\mu$m are distinguishable in the low-resolution spectra (R=10). For the higher-resolution (R=100) unbinned data, the error size is comparable to the transit depth above 9 $\mu$m so only the quartz feature is detectable. Corundum is only distinguishable for large grains in the low resolution spectra because of the shape at shorter wavelengths. The 13 $\mu$m corundum feature is only a 2-3 $\sigma$ detection for either grain size.}{Spectral features longward of 9 $\mu$m are only detectable in the binned spectra. }
}
\label{fig:spectra}
\end{figure*}

To simulate planet dust tail transits, we assume an optically-thin dust cloud with a constant opacity occulting the star. This is an unrealistic simplification; the actual dust tails will be expected to have decreasing opacity with increasing distance from the planet, as well as a decreasing grain size with increasing distance. 
However, the deepest part of the transit occurs when \replaced{the densest region of the cloud, which consists of a small area around the planet,}{the dense region of the dust tail that is slightly leading the planet} is occulting, so the effects from the changes in opacity and grain size should be small.
We also neglect limb darkening and assume the cloud is small compared to the star. The wavelength-dependent transit depth is 
\begin{equation}
\delta(\lambda) = \frac{N\, \bar{\sigma}(\lambda)_\mathrm{eff}}{\pi R_\star^2}
\end{equation}
where $N$ is the number of dust grains \added{in the cloud} and $\pi R_\star^2$ is the \replaced{cross section of the star}{stellar surface}. We scale $N$ to the transit depth in the visible.

\subsection{Results}\label{sec:JWST}

We calculate spectra using K2-22 as an example, scaling $N$ so that the transit depth is the average transit depth for K2-22 (0.55\%) in the \textit{Kepler} band. To estimate the errors, we use the \textit{JWST} transit error simulation code \textsc{PandExo} \citep{Batahla2017} with the K2-22 system properties listed in table \ref{tab:sysprop} and the \textit{JWST} MIRI Low Resolution Spectrograph (LRS) on slitless mode.  We assume a conservative noise floor of 50 ppm and calculate the results for a single transit with equal in-transit and out-of-transit integration times. We calculate the transmission spectra for a cloud of a single composition, shown in Fig.~\ref{fig:spectra} for large grain size ($r_\mathrm{eff}=1.0\,\mu$m) and small grain size distributions (0.1 $\mu$m) in the upper and lower plots, respectively. The left plots show the unbinned data at a resolution of $R=100$ with every 35th error bar (reduced for clarity) and the right plots display the data binned to $R=10$. %The right plots also include random noise in the data as estimated by \textsc{PandExo}. %The error calculations are explained below. 

The carbon and iron dust grains produce featureless spectra that are distinguished only by their relative depths at shorter wavelengths. Corundum has a single wide peak beginning near 12 $\mu$m with a small transit depth between 8 and 10 $\mu$m that distinguishes it from the other dust species. Quartz has a peak near 8.5 $\mu$m which is distinguishable for both grain size distributions. For the silicates, fayalite and forsterite have peaks at near 10 $\mu$m that are far enough apart to be distinguished even in a low-resolution spectrum. Enstatite and fayalite both have peaks near 10.5 $\mu$m, making them indistinguishable from this feature alone, but have distinct absorption properties between 6 and 8 $\mu$m for larger grains. However, for smaller grains, this degeneracy would only be distinguishable for resolutions of $R>10$. %For large grains, these peaks are preceded by a drop in transit depth. The breadth of this dip in transit depth can be used to distinguish enstatite and quartz. 
For each of the species with peaks, the depth is similar to that in the visible to about double the depth, \replaced{depending on the grain size.}{although the size of the features is sensitive to the grain size.}

%\begin{sidewaysfigure*}
\begin{figure*}
\centering
\mbox{
\includegraphics[width=5.25in]{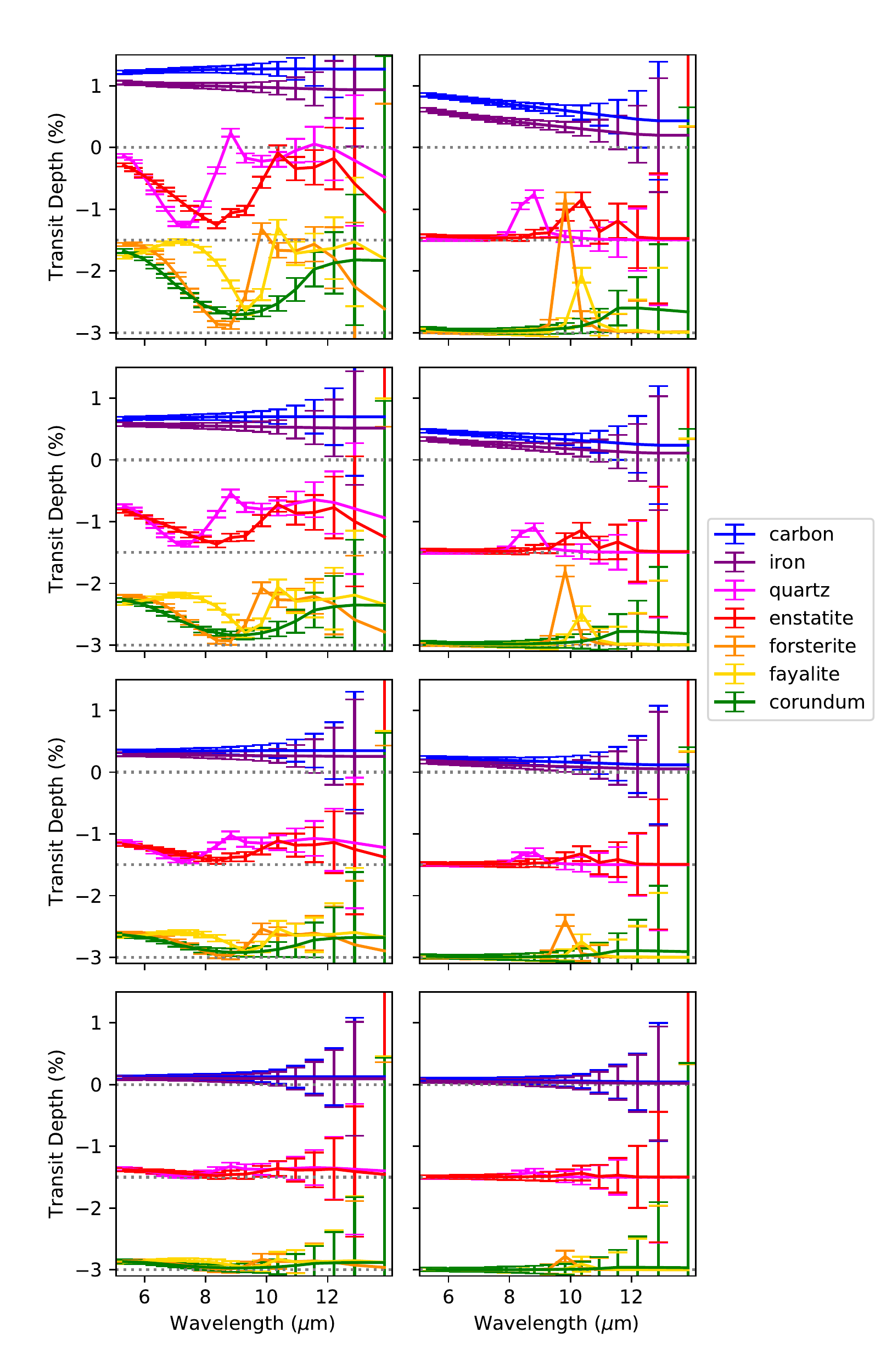}
}
\caption{Low-resolution (R=10) model spectra of a single transit for expected dust species for 1\%, 0.55\%, 0.275\%, and 0.1\% dips in the Kepler bandpass, plotted in order from top to bottom. The left plots are for the large grain size distribution ($r_\mathrm{eff}=1.0\,\mu$m) and the right plots are small grains ($r_\mathrm{eff}=0.1\,\mu$m). The same colors as in Fig. \ref{fig:spectra} mark the dust species and are offset in the same manner. Some features remain detectable down to a 0.275\% transit. The 13 $\mu$m corundum features remains undetectable for small grains at a 1\% transit depth and marginally detectable for large grains.}
\label{fig:sizecom}
\end{figure*}
%\end{sidewaysfigure*}

% For unbinned MIRI data, the error is too large to detect features beyond $\sim$10 $\mu$m. Binning down to R=10 reduces the SNR enough for a several sigma detection of fayalite, forsterite, and enstatite features, particularly with larger particles. 
%\added{Size has a large impact on the spectra, but for both distributions, the silicate features near 9 $\mu$m are distinguishable in the low-resolution spectra (R=10).}
Because these features are quite broad, low-resolution (R$\approx$10) spectra are sufficient to distinguish them, which allows us to bin the MIRI spectra enough to achieve sufficient SNR to detect them in the case of large grain size distribution.  \added{For the high-resolution (R=100) unbinned data, the error size is comparable to the transit depth longward of 9 $\mu$m so only the quartz feature is detectable.} The large grains broaden spectral features and create a small amount of background extinction \added{that is weakly wavelength-dependent, decreasing with increasing wavelength particularly between $6-8\,\mu$m}. Both of these effects in general create deeper spectral features that are easier to detect. Small grains have deeper spectral features but also thinner features which reduce in size quickly when the data are binned. \added{A higher-resolution binning would be preferable if the grains are small since it would make the features more prominent.}
The enstatite and fayalite features only have a marginally-detectable feature for the small grain size distribution while quartz and forsterite would have strong detections. For corundum, the 13 $\mu$m feature would only be a weak (1-2 $\sigma$) detection for either grain size, but if the grains are large, \replaced{the width of the dip in depth can be used to distinguish it}{the relative absorption between 6 and 8 $\mu$m and the lack of a feature at 10 $\mu$m make corundum distinguishable}. 

In Fig. \ref{fig:sizecom}, we examine a range of transit depths observed by \textit{Kepler} for this system. As expected, the features are larger with a larger transit, but even at 1\% the corundum feature is only marginally significant for small grains. At this larger transit depth, the breadth of the enstatite feature \added{at 10.5 $\mu$m} compared to fayalite becomes marginally detectable. For half the average depth (0.275\%), the silicate features are still detectable for large grains but only forsterite is detectable for small grains. At 0.1\%, no features for either grain size distribution have detectable compositional signatures. \added{The slopes of the spectra between 6-8 $\mu$m are still significantly different for some of the species, but the slope is degenerate with size in this wavelength region as the extinction cross-section is still dominated by the scattering cross-section. Only weak compositional constraints could be made at this small transit depth.} For a realistic cloud with multiple dust species, these features will be blended so the minimum visible transit depth required for detection of silicates is at least 0.3\%.

%\subsection{Signal-to-Noise Calculations for {\it JWST} Observations of Known Disintegrating Planets}
%\input{SNEst.tex}

% \subsection{Objectives and Proposed Work}
% We propose a combined {\em observational} and {\em theoretical modeling} effort to study disintegrating planets with the highest signal-to-noise ratios (SNR). 
% To make definite progress on both fronts, our work has the following components.

% \vspace{-0.02in}
% \subsubsection{Ground-based observations \label{sec:ground_obs}}
% 	\input{Ming_Jason/ground_obs_Ming_Jason.tex}

% \subsubsection{Dynamical models of ejected material}
% 	\input{Daniel/Dynamic_modeling_Daniel.tex}

% \subsubsection{Interior composition and mineralogy} 
% 	\input{Steve_Casey/Mineralogy_modeling_Steve_Casey.tex}

% \subsubsection{Preparation for JWST}
% 	\input{Ming_Jason/JWST_pilot_Ming_Jason_Steve.tex}

\section{Conclusions}\label{sec:conclusions}
We have calculated the detectability by infrared observations, in particular using {\it JWST}, of different mineral condensates in the tails of disintegrating planets, and encourage such observations.
The mineralogy of the dust will give us the first direct measurements of the chemical composition of the interiors of exoplanets and this can be be done with the capabilities of \textit{JWST}. Low-resolution transmission spectra allow for sufficiently high SNR in a single transit for a strong detection of any ferromagnesium silicate features that may exist in the dust tails of these planets if the transit depth in the \textit{Kepler} bandpass is at least 0.3\%. Since the average depth for K2-22b is 0.55\%, we should be able to observe at least one deep enough transit in a relatively low number of observations. With concurrent observations in the visible, several spectra may be combined to improve SNR despite the stochastic nature of the transit depths. If KIC 12557548b becomes more active again, then that system could also be characterized by \textit{JWST}, but KOI 2700b transit depths are currently too small.
\deleted{First and foremost, detection of these species would confirm the ``disintegrating planet" model for these systems. Dust condensed from material from a planet's interior would have a different composition than that from other sources such as comets. 
While cometary dust contains olivines and pyroxenes, it also has carbonaceous dust and polycyclic aromatic hydrocarbons (PAHs), and comets are rich in volatiles that would no longer be present on a magma ocean planet. In the wavelength region studied here, carbonaceous dust is indistinguishable from iron dust but PAHs have spectral features in the mid-IR.
These minerals are not expected from evaporated planetary material, so their detection would be a test for the disintegrating planet model.}

\replaced{Second, assuming a planetary origin,}{If the dust is condensing from rocky vapor from a molten surface,} the relative abundances in the dust species will allow us to determine whether it is crustal, mantle, or core material that is escaping. We expect the crust would produce more Al-bearing minerals than interior material, while the core would produce primarily iron-bearing minerals. Since the crustal layer is thin, crustal material would disintegrate in a short period of time compared to the mantle. Detection of crustal material may be less likely, but if discovered it would constrain the time for which the planet has been disintegrating. \added{If the dusty affluents are from volcanic plumes, then the presence of volatile elements such as carbon may be an indicator.}

Assuming mantle material is escaping and the dust is representative of the mantle composition, determining the abundances of the Mg-, Si-, and Fe-bearing minerals can be used to derive the Mg/Si and Fe/Si ratios of the dust which then can constrain these ratios in the planet's interior. Core size probably correlates directly with Fe/Si, and many important geophysical properties of the mantle --- water storage capacity, thermal conductivity, rheology, etc. --- depend on the mineralogy of the mantle. 
\deleted{The dynamics of the mantle may be sensitive to Mg/Si ratios, which may  determine whether plate tectonics can be driven, which in turn affects planet habitability.} 
If more disintegrating planets are found in future missions such as \textit{TESS}, we can use these planets to test how good of a proxy the stellar composition is for the composition of the planet, although precise stellar composition observations are also required to test this.

\acknowledgements
The authors thank Brad Foley for useful discussions, Daniel Jontof-Hutter for comments, and Natasha Batalha for assistance with the Pandexo code.

This work was partially supported by funding from the Center for Exoplanets and Habitable Worlds, which is supported by the
 Pennsylvania State University, the Eberly College of Science, and the Pennsylvania Space Grant Consortium.
E.H.L.B.'s research was supported by an appointment to the NASA Postdoctoral Program
with the Nexus for Exoplanet System Science, administered by Universities Space Research
Association under contract with NASA.

\bibliography{references}


\end{document}